\documentclass{ws-procs975x65}

\begin{document}

\title{INTER-UNIVERSAL QUANTUM ENTANGLEMENT}

\author{S. J. ROBLES-P\'{E}REZ}

\address{F\'{\i}sica Te\'orica, Universidad del Pa\'{\i}s Vasco, Apartado 644, 48080, Bilbao (SPAIN).}

\author{P. F. GONZ\'{A}LEZ-D\'{I}AZ}

\address{Colina de los Chopos, Centro de F\'{\i}sica ``Miguel Catal\'{a}n'', Instituto de Matem\'{a}ticas y F\'{\i}sica Fundamental, Consejo Superior de Investigaciones Cient\'{\i}ficas, \\
Serrano 121, 28006, Madrid (SPAIN) and \\ 
Estaci\'{o}n Ecol\'{o}gica de Biocosmolog\'{\i}a, Pedro de Alvarado, 14, 06411-Medell\'{\i}n, (SPAIN).}

\begin{abstract}
The boundary conditions to be imposed on the quantum state of the whole multiverse could be such that the universes would be created in entangled pairs. Then, inter-universal entanglement would provide us with a vacuum energy for each single universe that might be fitted with observational data, making testable not only the multiverse proposal but also the boundary conditions of the multiverse. Furthermore, the second law of the entanglement thermodynamics would enhance the expansion of the single universes.
\end{abstract}


\bodymatter

\section{Introduction}\label{sec1}

The boundary condition that should be imposed on the quantum state of the whole multiverse could imply the existence of quantum correlations among the states of different universes of the multiverse. In particular, it is plausible to consider that some of the universes of the multiverse could be created in entangled pairs.

The existence of quantum correlations among the universes of the multiverse would make that, even being these classically disconnected, the properties of a single universe might be influenced by the presence of other universes of the multiverse whose quantum states were correlated. First, the energy of entanglement between the state of two universes would supply each single universe with a contribution to their energy density that should leave testable effects on their evolution, at least in principle. Second, the entropy of entanglement, i.e. the entropy that corresponds to the reduced density matrix that represents one single universe of the entangled pair, would  provide us with an arrow of time for each single universe and it might have some influence on the growth of cosmic structures and other processes that are customary involved in the cosmological arrow of time provided that the thermodynamics of entanglement is a quantum generalization of the classical formulation of thermodynamics\cite{Brandao2008}.

\section{Quantum state of the multiverse}

Let us consider the multiverse that corresponds to a multiply-connected manifold formed by a set of simply-connected regions of the space-time which will  be called universes hereafter. Each single universe is classically disconnected from the rest of universes, which would thus be disregarded as being physically admissible for an observer inhabiting a singled out universe. However, they should be taken into account if quantum correlations do exist in the composite state of the whole set.

For each single universe we shall consider a homogeneous and isotropic space-time filled with a slow-varying field, $\varphi$, that will eventually give rise to the matter content of the universe. The Wheeler-deWitt equation can then be written as\cite{RP2012b} 
\begin{equation}\label{WDW}
\ddot{\phi} + \frac{\dot{\mathcal{M}}(a)}{\mathcal{M}(a)} \dot{\phi} + \omega^2(a,\varphi) \phi = 0 ,
\end{equation}
where, $\mathcal{M}(a) = a$ and $\omega(a,\varphi) = \frac{a}{\hbar}\sqrt{a^2 V(\varphi) - 1}$, with $V(\varphi)$ being the potential of the scalar field, $\phi\equiv \phi(a,\varphi)$ is the wave function of the universe and the over-dot means the derivative with respect to the scale factor, $a$.

Following the analogy between Eq. (\ref{WDW}) and the equation of a harmonic oscillator, the Hamiltonian for which the Heisenberg equations give rise to the Wheeler-deWitt equation (\ref{WDW}) is given by the Hamiltonian of a harmonic oscillator with scale factor dependent mass, $\mathcal{M}(a)$, and frequency, $\omega(a,\varphi)$. Now, the boundary condition that the number of universes of the multiverse does not depend on the value of the scale factor of a particular single universe restricts the possible representations that can be considered to the set of invariant representations\cite{Kim2001}. However, in terms of the creation and annihilation operators of a given invariant representation, $\hat{b}_I^\dag$ and $\hat{b}_I$, respectively, the Hamiltonian of the harmonic oscillator (\ref{WDW}) turns out to be
\begin{equation}\label{Hamiltonian}
\hat{H} = \hbar \left\{ \beta_+ (\hat{b}_I^\dag)^2 + \beta_- \, \hat{b}_I^2 + \beta_0 \left( \hat{b}_I^\dag \hat{b}_I + \frac{1}{2} \right)  \right\} ,
\end{equation}
for some functions $\beta_\pm(a,\varphi)$ and $\beta_0(a,\varphi)$ whose values depend on the particular representation chosen\cite{Kim2001}. The Hamiltonian (\ref{Hamiltonian}) is formally similar to the Hamiltonian of the degenerated parametric amplifier which is related in quantum optics with processes that involve the creation of an entangled pair of photons \cite{Scully1997}. Similarly, we can interpret that the quadratic operators in Eq. (\ref{Hamiltonian}) are associated to the creation of entangled pairs of universes in the quantum multiverse.

\section{Thermodynamics of inter-universal entanglement}

Let us consider an entangled pair of universes whose composite quantum state is given by the two-mode squeezed state\cite{RP2010}
\begin{equation}
\rho = \mathcal{U}(a,\varphi) |0\rangle\langle 0 | \mathcal{U}^\dag(a,\varphi) ,
\end{equation}
where, $\mathcal{U}\equiv e^{r(a,\varphi) \hat{b}_1^\dag \hat{b}_2^\dag - r^*(a,\varphi) \hat{b}_1 \hat{b}_2}$, with $r(a,\varphi)$ being the parameter of squeezing and $\hat{b}_{(1,2)}^\dag$ and $\hat{b}_{(1,2)}$ being the ladder operators for the corresponding Fock spaces of each single universe, labelled by $1$ and $2$, respectively. The reduced density matrix that represents the quantum state of each single universe is obtained by tracing out from $\rho$ the degrees of freedom of the partner universe. It turns out to describe a thermal state given by\cite{RP2010, RP2012b}
\begin{equation}\label{ts}
\rho_1 \equiv {\rm Tr}_2 \rho = 2 \sinh\frac{\omega}{2 T} \sum_{N=0}^\infty e^{-\frac{\omega}{T} (N+\frac{1}{2})}  |N\rangle\langle N| .
\end{equation}
The two universes of the entangled pair evolve then in thermal equilibrium with respect to each other with a temperature, $T \equiv T(a,\varphi) \equiv -\frac{\omega(a,\varphi)}{2 \ln\Gamma(a,\varphi)}>0$, that depends on the value of the scale factor. It is straightforward to compute the thermodynamical magnitudes\cite{Alicki2004} associated to the thermal state (\ref{ts}). The entropy of entanglement, $S_{ent}$, given as usual by the von-Neumann formula, turns out to be a decreasing function with respect to an increasing value of the scale factor. However, the second principle of thermodynamics is still satisfied because the evolution of each single universe is not adiabatic, in the quantum informational sense, and the heat decreases as the universe expands. More exactly, the production of entropy\cite{Alicki2004} is zero and
\begin{equation}
dS_{ent} = \frac{\delta Q}{T} < 0 .
\end{equation}

\section{Conclusions}

Within the general picture of a multiverse made up of entangled pairs of universes, the entropy of the multiverse would be zero provided that its quantum state is given by a pure state. We might say that concepts like \emph{time} and \emph{evolution} really make sense within single universes. For these, inter-universal entanglement provides an arrow of time\cite{RP2012}. The entropy of entanglement decreases with respect to an increasing value of the scale factor, satisfying however: i) the second principle of (classical) thermodynamics, and ii) the second principle of entanglement thermodynamics\cite{Plenio1998} provided that the universe expands decreasing thus the amount of entanglement between the universes.

The energy of entanglement, $E_{ent}\equiv Q$, would contribute to the total energy of a single universe and, therefore, it would presumably leave some trace on the evolution of the universe. That would make testable the whole multiverse proposal. However, it is expected that such an effect is subdominant except, may be, during some special stages of the universe.

\section*{Acknowledgments}
This work was supported by the Basque Country Government project IT-221-07.

\bibliographystyle{ws-procs975x65}
\bibliography{bibliography}

\begin{thebibliography}{1}

\bibitem{Brandao2008}
F.~G. S.~L. Brandao and M.~B. Plenio, {\em Nature Physics} {\bf 4}, 873 (2008).

\bibitem{RP2012b}
S.~J. Robles-P{\'e}rez, {\em Open questions in cosmology} (InTech, 2012),
  ch.~Inter-universal entanglement.

\bibitem{Kim2001}
S.~P. Kim and D.~N. Page, {\em Phys. Rev. A} {\bf 64}, p. 012104 (2001).

\bibitem{Scully1997}
M.~O. Scully and M.~S. Zubairy, {\em Quantum optics} (Cambridge University
  Press, Cambridge, UK, 1997).

\bibitem{RP2010}
S.~Robles-P{\'e}rez and P.~F. Gonz{\'a}lez-D{\'\i}az, {\em Phys. Rev. D} {\bf
  81}, p. 083529 (2010).

\bibitem{Alicki2004}
R.~Alicki {\em et~al.}, {\em Open Syst. Inf. Dyn.} {\bf 11}, 205 (2004).

\bibitem{RP2012}
S.~Robles-P{\'e}rez, {\em (submitted)}  (2012).

\bibitem{Plenio1998}
M.~B. Plenio and V.~Vedral, {\em Comtemp. Phys.} {\bf 39}, 431 (1998).

\end{thebibliography}

\end{document}